\documentclass[extra, mreferee]{gji} 
\usepackage[urlcolor=blue,citecolor=black,linkcolor=black]{hyperref}
\usepackage{graphicx}
\usepackage{amsmath}
\usepackage{color}

\newcommand{\GJIFooterText}{%
  This work is published at \textit{Geophys.\ J.\ Int.}. %
  Please visit the original publication at %
  \url{https://doi.org/10.1093/gji/ggaf397}.%
}

\makeatletter
\newcommand{\InstallGJIFooter}{%
  \def\@oddfoot{\footnotesize
    \hspace*{0pt}\parbox{\textwidth}{\centering \GJIFooterText}}%
  \def\@evenfoot{\footnotesize
    \hspace*{0pt}\parbox{\textwidth}{\centering \GJIFooterText}}%
}
\makeatother

\title{DASPack: Controlled Data Compression for Distributed Acoustic Sensing}
\author[A. Seguí, A. Ugalde, A. Fichtner,  S. Ventosa and R. Morros ]
  {Aleix Seguí$^{1}$, Arantza Ugalde$^{2}$, Andreas Fichtner$^{1}$, Sergi Ventosa$^{2}$, \and Josep Ramon Morros$^{3}$ \\
   $^{1}$ ETH Zürich, Zürich, Switzerland \\
   $^{2}$ Institute of Marine Sciences, CSIC, Barcelona, Spain \\
   $^{3}$ Universitat Politècnica de Catalunya (UPC), Barcelona, Spain}
\date{Accepted 2025 October 07. Received 2025 October 07; in original form 2025 July 17}
\pagerange{\pageref{firstpage}--\pageref{lastpage}}
\pubyear{{\the\year{}}}

\begin{document}

\InstallGJIFooter

\label{firstpage}

\maketitle

\begin{summary}
We present DASPack, a high-performance, open-source compression tool specifically designed for distributed acoustic sensing (DAS) data. As DAS becomes a key technology for real-time, high-density, and long-range monitoring in fields such as geophysics, infrastructure surveillance, and environmental sensing, the volume of collected data is rapidly increasing. Large-scale DAS deployments already generate hundreds of terabytes and are expected to increase in the coming years, making long-term storage a major challenge. Despite this urgent need, few compression methods have proven to be both practical and scalable in real-world scenarios. DASPack is a fully operational solution that consistently outperforms existing techniques for DAS data. It enables both controlled lossy and lossless compression by allowing users to choose the maximum absolute difference per datum between the original and compressed data. The compression pipeline combines wavelet transforms, linear predictive coding, and entropy coding to optimise efficiency. Our method achieves up to $3\times$ file size reductions for strain and strain rate data in lossless mode across diverse datasets. In lossy mode, compression improves to $6\times$ with near-perfect signal fidelity, and up to $10\times$ is reached with acceptable signal degradation. It delivers fast throughput (100–200 MB s\(^{-1}\) using a single-thread and up to 750 MB s\(^{-1}\) using 8-threads), enabling real-time deployment even under high data rates. We validated its performance on 15 datasets from a variety of acquisition environments, demonstrating its speed, robustness, and broad applicability. DASPack provides a practical foundation for long-term, sustainable DAS data management in large-scale monitoring networks.

\end{summary}


\section{Introduction}

Distributed acoustic sensing (DAS), turning standard fibre-optic cables into dense arrays of strain and strain rate sensors, is a widely used technique for high-density and large-scale monitoring \citep[e.g.,][]{Hartog_2017,Zhan_2020,Lindsey_2021}. DAS has grown in popularity across geophysics, seismology, and engineering. While early applications included infrastructure monitoring \citep[e.g.,][]{Owen_2012,Hill_2015}, DAS has been successfully applied in borehole seismic imaging \citep[e.g.,][]{Mateeva_2014,Daley_2016} and near-surface exploration and monitoring \citep[e.g.,][]{Daley_2013,Martin_2017,Tuinstra_2024}. More recently, it has opened new opportunities for seismic observation in settings where conventional instruments are difficult to deploy, such as urban areas \citep[e.g.,][]{Lindsey_2017,Biondi_2017,Ajo_2019,Spica_2020,Smolinski_2024}, ocean floors \citep[e.g.,][]{Lior_2021,ugalde2021,Igel_2024}, active volcanoes \citep[e.g.,][]{Jousset_2022,Klaasen_2022,Klaasen_2023}, and the cryosphere \citep[e.g.,][]{Booth_2020,Walter_2020,Hudson_2021,Fichtner_2025}. DAS has also enabled the detection of dynamic processes in unstable slopes \citep[e.g.,][]{Ouellet2024,Kang_2024}.

Yet, the scientific gains come with a concerning data management challenge: for example, a single interrogator recording 5,000 channels at 1 kHz using 4-byte floating-points produces nearly 1.5 TB per day, and multi-interrogator campaigns can exceed hundreds of terabytes per month; a number that is likely to increase in the near future. The large data generation rates also entail impractical data transfer times. In many cases, users resort to physically transporting storage disks, which becomes increasingly unfeasible as the projects expand. Long-term archiving initiatives are similarly limited and face expensive storage challenges \citep{NSF_SAGE_Data}. 

We propose using data compression as an effective solution to address these issues. In particular, we study a controlled data compression approach,  where data can be reconstructed after rounding at a desired decimal value. While lossy compression can achieve greater size reduction, it may discard critical information, limiting the value of the data for precise scientific analyses. Hence, a controlled compression method, including lossless mode, is imperative for applications that demand complete data integrity \citep{Segui2024,Bozzi2025}. 

Existing approaches often compress each channel independently. \citet{dong22} proposed a lossless integer compression method, H5TurboPFor, that provides both speed and efficiency by combining a preprocessing step with a general one-dimensional integer compressor. \citet{wu23} employed a 2-D redundancy exploitation strategy using one-step channel differences, achieving approximately a 50\% reduction in data size. General-purpose dictionary-based compression algorithms, such as GZIP and Zstd, are also widely used, but are known to typically offer lower compression efficiency \citep{dong22}. ZFP \citep{lindstrom14} is another method used in the DAS community, which focuses on floating points and supports both lossless and lossy compression of multidimensional arrays, making it suitable for DAS applications \citep{issah24}. We evaluate their performance in our results section.

Given that DAS data naturally forms a 2-D structure (time and distance, where each channel corresponds to a specific position along the fibre), image compression approaches are also appealing. However, their direct application is challenging: the time and distance domains in DAS data exhibit highly unbalanced correlations, unlike the patterns found in natural images. Moreover, compression speeds optimised for image data can be too slow for DAS applications, which typically involve much larger data volumes. In the previous example, 5,000 channels at 1 kHz can generate data at a rate about 20~MB~s$^{-1}$. Even so, methods like JPEG2000 \citep{taubman2012}, based on wavelet transforms and advanced entropy coding, have been explored in various fields for 2-D data compression. Additionally, wavelets have been known to be effective for compression of seismic data \citep{vassiliou97, Sebai2024}. Therefore, we include JPEG2000 in our comparative study.

Recent studies have explored deep-learning-based compression for DAS. \citet{yiyu2024} used a lightweight model to achieve reasonable wavefield reconstruction from $5\times$ subsampling. \citet{chen2025} achieved up to $8\times$ lossy compression using a vision transformer on 10-second DAS tiles, which can be further enhanced through downsampling techniques to achieve greater reductions in data size. However, learned compression methods currently face significant limitations. These include a lack of generalisability across datasets, inability to guarantee explicit error bounds, and the tendency to introduce data-driven reconstruction artifacts that can inadvertently regenerate patterns present in the training data. Moreover, such models are computationally intensive, often requiring resources that prevent real-time deployment, especially on CPU-based systems. Only lightweight variants with moderate compression power can operate at reasonable speed. As a result, these approaches remain impractical for many real-world scenarios in their current form. 

In this paper, we present DASPack, a method that combines a user-defined data degradation step with a lossless compression pipeline to achieve effective compression. The lossless compression pipeline begins with wavelet decomposition, which separates the signals in frequency and wavenumber (i.e., spatial frequency) components. A linear prediction filter is then applied to exploit redundancy in both the temporal and spatial dimensions. This results in a decorrelated, low-variance signal that is well-suited for effective entropy coding, which encodes the data into the shortest possible bit sequences based on symbol probability distribution. All steps in the pipeline are fully reversible, allowing to compress losslessly if desired.

Our approach seeks to boost compression performance compared to the most widely used generalist methods and specialised alternatives, while maintaining fast processing speed, to be able to handle large DAS data volumes seamlessly. We provide a fast, real-time, and fully open-source implementation of this pipeline under the name DASPack. The software is developed in Rust \citep{rustbook}, a programming language designed to build reliable and efficient software. This choice allows for fine grain control over computational resources while ensuring memory safety, making DASPack both fast and robust. Furthermore, DASPack can be accessed directly from Python, and the data is stored in HDF5 format.

\begin{table*}
\caption{Datasets used in the study. An index (Id.) is assigned for referencing through the text and figures. Datasets 1-8 recorded nanostrain rate ($\text{nm} ~\text{m}^{-1}~\text{s}^{-1}$) data while datasets 9-15 recorded nanostrain ($\text{nm} ~\text{m}^{-1}$). For published datasets, the corresponding journal publication is provided in the Reference column. Three datasets (Ids. 4, 7 and 12) do not have a journal publication: Id. 4 was acquired in November 2023 during a volcanic crisis in Iceland, using a $\sim$8 km long telecommunication cable extending from the Blue Lagoon to the town of Grindavík. Id. 7 was acquired in April 2022 using a 1 km long fibre-optic cable deployed in an old irrigation tunnel near Guía de Isora, which extends into the mountain towards the peak of El Teide. Id. 12 was acquired between February and May 2022 along $\sim$60 km of the Balalink submarine telecommunications cable, connecting Valencia with Palma de Mallorca.}
\begin{minipage}{7in}
\footnotesize 
\label{tab:datasets}
\begin{tabular}{clllrrr}
\textbf{Id.} & \textbf{Location}            & \textbf{Environment} & \textbf{Reference} & \textbf{\shortstack{Sampling \\ Rate [Hz]}} & \textbf{\shortstack{Channel \\ Spacing \\ (Gauge \\ Length) [m]}}  & \textbf{\shortstack{Num. of \\ Channels}} \\ \hline
1  & VdlS (Switzerland)           & Landslides           &  \citet{aichele2025}                  & 400    & 2 (8)   & 2176 \\
2  & Athens (Greece)              & Urban                &  \citet{tian2022}                  & 200    & 2 (8)   & 7424 \\
3  & Rhonegletscher (Switz.)      & Glacier              & \citet{walter2020}                   & 500    & 4 (8)   & 576  \\
4  & Blue Lagoon (Iceland)        & Volcanic             &   --                 & 500    & 4 (8)   & 512  \\
5  & Susch (Switzerland)          & Landslides           & \citet{edme2023}                   & 100    & 2 (8)   & 5120 \\
6  & Fagradalsfjall (Iceland)     & Volcanic             &  \citet{klaasen21b}                  & 400    & 2 (8)   & 2496 \\
7  & Tenerife (Spain)             & Volcanic             &  --                  & 200    & 2 (8)   & 1280 \\
8  & Zurisee (Switzerland)        & Lake                 & \citet{rychen2023}                   & 10\,000 & 1 (8)  & 2496 \\ 
9  & Mt.\ Meager (Canada)         & Volcanic             & \citet{klaasen2021} & 1\,000  & 8 (7)   & 375  \\ 
10 & Alboran Sea (Spain)          & Submarine            & \citet{ugalde2023b} & 100     & 10 (10) & 5984 \\ 
11 & Gran Canaria (Spain)         & Submarine            & \citet{ugalde2023a} & 50      & 10 (10) & 7008 \\ 
12 & Valencia (Spain)             & Submarine            & --                  & 250     & 10 (10) & 5983 \\ 
13 & Vinaroz (Spain)              & Submarine            & \citet{villasenor2023} & 100   & 10 (10) & 3488 \\ 
14 & Tenerife (Spain)             & Volcanic             & \citet{ugalde2021} & 100   & 10 (10) & 2176 \\ 
15 & La Palma (Spain)             & Volcanic             & \citet{villasenor21}  & 100   & 10 (10) & 832  \\ \hline
\end{tabular}
\end{minipage}
\end{table*}

\section{Data}

Datasets from diverse sources were compiled for a thorough benchmarking of the compression methods. Recorded data types include strain (measured in nanostrain ($\text{n}\varepsilon$), equivalent to nanometres per metre or $\text{nm} ~\text{m}^{-1}$) and strain rate (measured in nanostrain per second ($\text{n}\varepsilon~\text{s}^{-1}$), equivalent to nanometres per metre per second or $\text{nm} ~\text{m}^{-1} ~\text{s}^{-1}$). Table \ref{tab:datasets} summarises the datasets used in this study. Datasets 1-8 were recorded using a Silixa iDAS interrogator, storing data with 16-bit integers. Dataset 9   was recorded with an OptaSense ODH3 unit storing 32-bit integers, while datasets 10-15 were collected using an Aragon Photonics HDAS interrogator and stored 32-bit floating-points. The datasets span a wide range of environments, including urban settings, landslide-prone areas, glaciers, volcanic zones, lakes, and offshore regions. This diversity ensures that DASPack is evaluated under realistic and varied signal conditions. The datasets also differ in key acquisition parameters such as signal type (strain and strain rate), data type (integers and floating-points), sampling rate (ranging from 50 Hz to 10,000 Hz), and channel spacing. The number of channels defines the distance extent and varies from a few hundred to above 7,000, allowing to test the scalability of the compression pipeline.

\section{Methods}

\subsection{DASPack Overview}

DASPack compresses DAS records in two decoupled stages that separate scientific requirements (how much error is acceptable) from engineering optimisation (how to store the data with minimal bits). The stages are:
\begin{enumerate}
    \item \emph{Controlled Degradation}.  Data amplitudes are first mapped from their raw representation (integers or floating points) to signed integers by rounding the data to a user-defined resolution. This mapping fixes the absolute reconstruction error so the user can choose from mathematically lossless storage to more aggressive, but explicitly quantified, data reduction. 
    \item \emph{Lossless Compression}.  The integer array is passed through a sequence of perfectly invertible transforms that progressively lower its entropy and a final coding step that creates an optimised binary file.  Each transform is exactly reversible, so the only potential information loss is the user-specified rounding at the first step. 
\end{enumerate}

\subsection{Controlled degradation}

\begin{figure*}
    \centering
    \includegraphics[width=\linewidth]{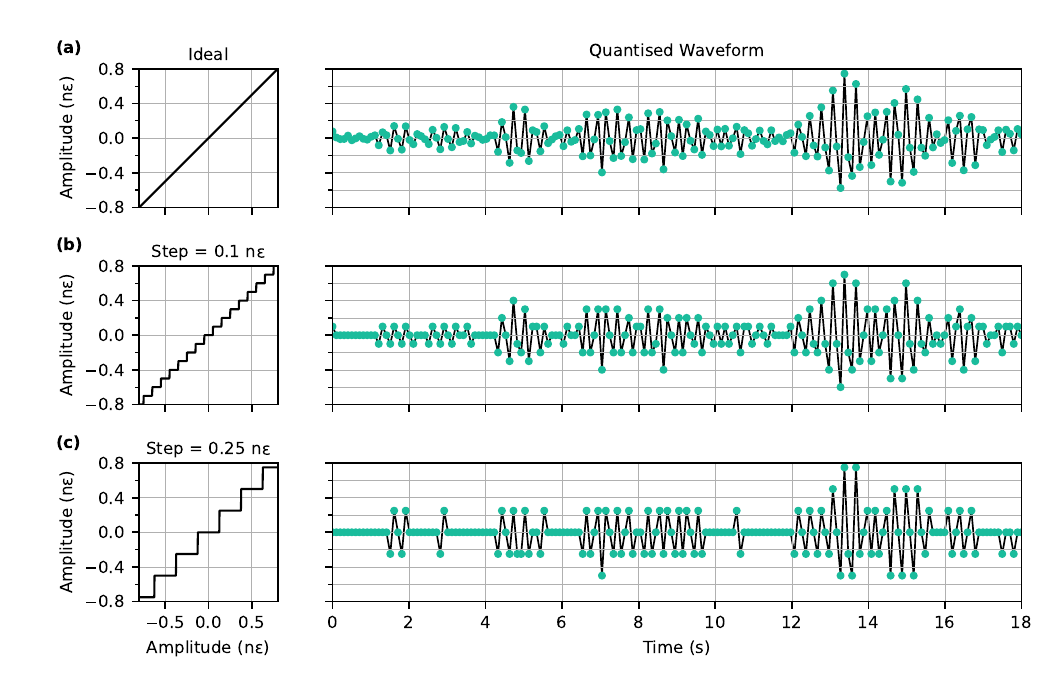}
    \caption{Controlled degradation of the amplitude in nanostrain ($\text{n}\varepsilon$) at different quantisation steps. The data points are represented as coloured dots, and are joined via a black line. The continuous, original signal is represented in (a), (b) shows a quantisation step of 0.1$~\text{n}\varepsilon$  and (c) is 0.25 $~\text{n}\varepsilon$. The waveform is extracted from dataset 10 (Alboran Sea). }
    \label{fig_discretisation}
\end{figure*}

Lossy compression techniques used in image and audio processing reduce data size by discarding perceptually insignificant information. These methods are tailored to human perception and often rely on subjective quality metrics. In scientific applications subjectivity is unsuitable, and the reference low seismic background noise model \citep{Peterson1993}, that could potentially be used in seismometer and accelerometer recordings as an alternative, is not appropriate for strain and strain rate DAS recordings. For this reason, DASPack adopts a principled, user-defined approach to degradation based on fixed-accuracy quantisation.

In DASPack, controlled degradation is implemented as a single, explicit rounding step prior to lossless compression. This fixed-accuracy approach maps the original data to signed integers, effectively redefining the resolution of the analogue-to-digital conversion post-acquisition,
\begin{equation}
    d_{\text{rounded}} = \text{round}\left(\frac{d_{\text{orig}}}{\Delta}\right) \cdot \Delta,
\end{equation}

where $d_{\text{orig}}$ is the raw datum and $\Delta$ is the chosen step size, leading to a maximum error introduced by compression of $\frac{\Delta}{2}$. This quantisation ensures that the reconstruction error is strictly bounded and interpretable. 

Unlike filtering or decimation, quantisation preserves the original signal bandwidth, maintaining both temporal and spatial resolution. For smaller $\Delta$, the rounding error tends to be uniform and uncorrelated across time and channels, resembling additive white noise \citep{Widrow_2008}. 

Fig. \ref{fig_discretisation} exemplifies the effect on an earthquake waveform. The degradation removes all signals that have an amplitude smaller than the quantisation step. We observe that, for this particular example, at $0.1 ~\text{n}\varepsilon$  the very low amplitude signal is diminished, and at a very agressive step of $0.25 ~\text{n}\varepsilon$, significant information is lost.

\begin{figure*}
    \centering
    \includegraphics[width=0.95\linewidth]{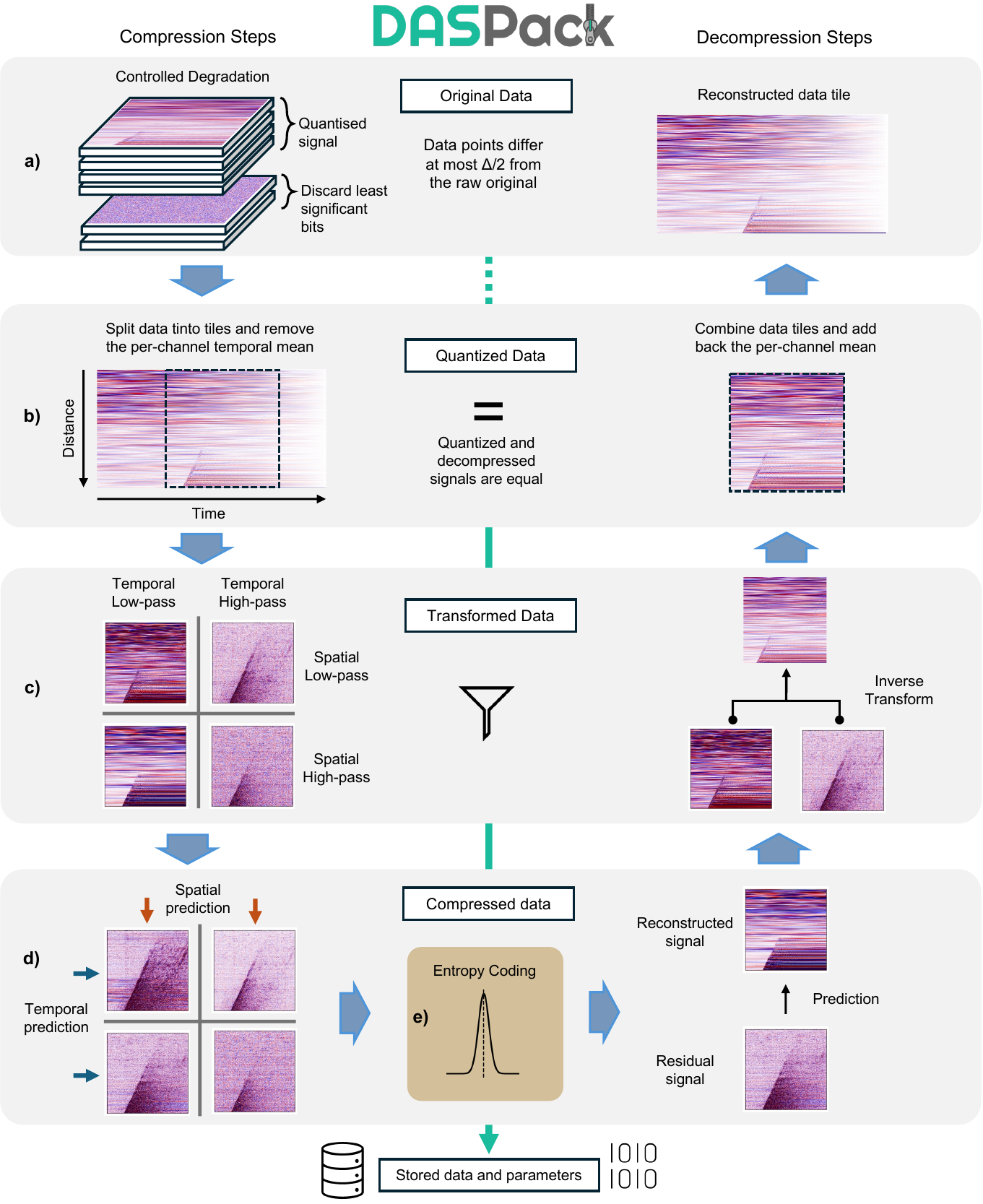}
    \caption{Compression and decompression pipelines. (a) Data is transformed to integers and optionally quantised via a user-defined quantisation step. (b) Data is first split into non-overlapping tiles that are processed separately and the temporal mean is removed per-channel. (c) A wavelet transform is optionally applied in both (time and space) dimensions. (d) A linear prediction filter is optionally applied with a fixed number of parameters in both dimensions in every wavelet subband. (e) Data is entropy coded to obtain a bit sequence. Step (a) is irreversible but tuneable; steps (b), (c), (d) and (e) are all reversible, ensuring lossless compression.    }
    \label{fig_main}
\end{figure*}

\subsection{Compression Pipeline}
\label{sec:pipe}

DASPack's main contribution is the compression pipeline that integrates several complementary techniques, each serving a different role to achieve effective compression. The presented method is designed for integer arithmetic and the degradation step is responsible for converting data into tractable integers. 

The proposed compression pipeline is depicted in Fig. \ref{fig_main}, which shows a reversible process, including both compression and decompression steps. The steps are illustrated using an example from a submarine fibre-optic cable from Vinaroz (dataset 13) that is deployed at very shallow depth (less than 60 m) and contains a seismic event of magnitude \textit{mbLg} 1.6 occurred on 20 July 2023 at 11:49:43 UTC. The stored data is also indicated, consisting of compressed 2-D arrays and side information, such as the predictive model coefficients. Below, we detail each step of the DASPack compression pipeline:
\begin{enumerate}
\renewcommand{\theenumi}{(\alph{enumi})}
    \item First, the raw data is converted to integers according to the desired absolute accuracy (Fig. \ref{fig_main}a) , defined as the maximum absolute difference between the compressed and original data. This step becomes irreversible (lossy) if the original data is in floating-point format, or if the quantisation step exceeds the resolution of the raw integer data. The same degradation is applied over all input data, avoiding discontinuities or jumps.
    \item Next, the data is divided into non-overlapping windows (or tiles), which are processed independently (Fig. \ref{fig_main}b). The optimal size of each tile depends on various data characteristics, such as temporal and spatial resolution or signal entropy. This design also allows for parallel processing of data. To improve the effectiveness of subsequent steps, the temporal mean is subtracted from each channel and stored as side information.
    \item Then, a discrete wavelet transform is optionally applied along both the time and distance dimensions (Fig. \ref{fig_main}c). The wavelet transform decomposes data in several frequency and wavenumber bands along the time and distance dimensions, respectively, without increasing the total number of samples and allowing afterward a perfect reconstruction of the original signal \citep{JACQUES_2011}. 
    \item A linear predictive coding (LPC) model is computed separately for each row (temporal dimension) and each column (spatial dimension) of the low-pass wavelet subbands (Fig. \ref{fig_main}d). The LPC prediction captures the dominant low-frequency components of the signal, which typically account for most of its energy. As a result, the residual (the difference between the prediction and the actual signal) has lower variance, improving the effectiveness of subsequent step (entropy coding). The optimal, discretised prediction coefficients are applied during compression and stored as side information to ensure exact reconstruction during decompression.  
    \item Finally, the residual signal is then converted into a compact bitstream using entropy coding and stored on disk, along with the associated optimal parameters (Fig. \ref{fig_main}e). The decompression algorithm reverses the above steps. Integer operations ensure a mathematically exact reconstruction up until degradation (back from (e) to (b)). The only difference between the decompressed data and the original is the controlled degradation in step (a).  
\end{enumerate}

Following this summary, we now describe the different parts in more detail.

\subsection{Discrete Wavelet Transform}

 A wavelet basis is a set of functions used to decompose signals into components that capture both frequency and location information \citep{stephane2009}. To enable perfect reconstruction of a signal from its components, a complete orthogonal (or biorthogonal) basis of wavelet functions is required. Discrete wavelets act a carefully designed filter bank that iteratively decompose a signal into low-frequency (approximation) and high-frequency (detail) components, without increasing the total number of samples. We employ the LGT 5/3 wavelet \citep{legall1988}, which is a biorthogonal wavelet commonly used in image compression, particularly in the JPEG2000 standard \citep{taubman2012}. This wavelet is implemented through the lifting scheme \citep{daubechies1998}, as done in JPEG2000, for computational efficiency and exact integer arithmetic.

\subsection{Linear Predictive Coding}

\begin{figure*}
    \centering
    \includegraphics[width=\linewidth]{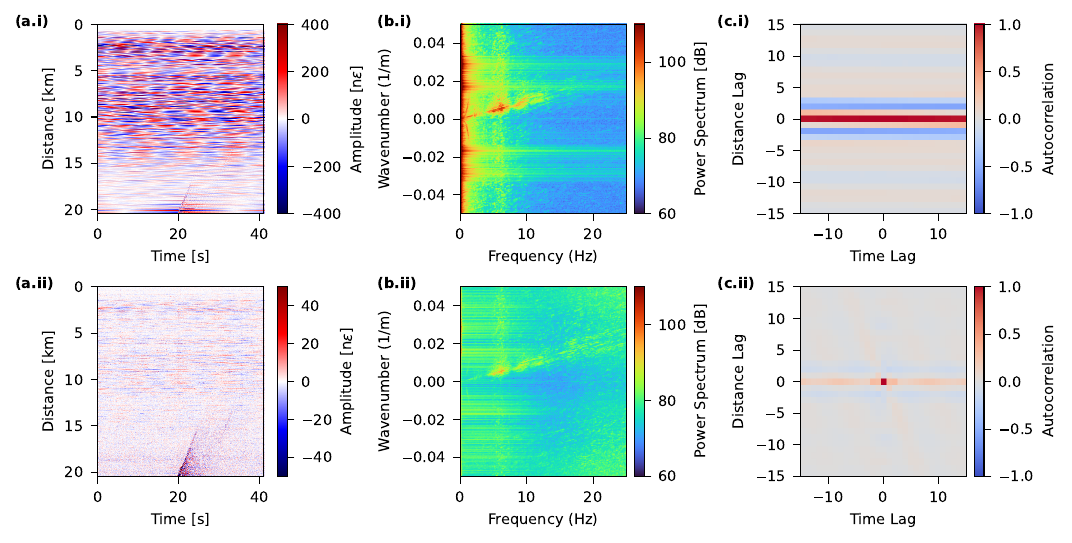}
    \caption{(i) Raw original and (ii) LPC-filtered data example, showing strain from the dataset 13 (Vinaroz), see Table \ref{tab:datasets}. Column (a) shows the strain values, (b) the power spectrum in the $f$-$k$ transform and (c) the 2-D autocorrelation function. }
    \label{fig_lpc}
\end{figure*}

\begin{figure*}
    \centering    \includegraphics[width=\linewidth]{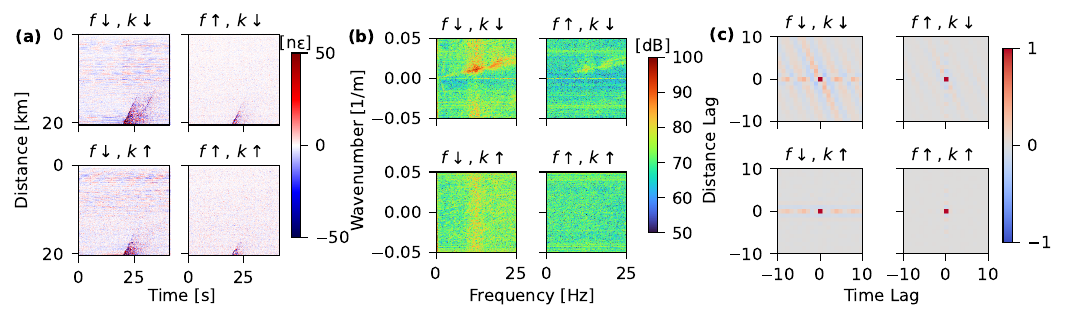}
    \caption{Example of wavelet-transformed and LPC-filtered strain data from dataset 13 (Vinaroz). For every subfigure, the four subbands obtained after the applying the wavelet transform in both dimensions are shown. Left plots correspond to frequency low-pass and right plots, to frequency high-pass. Top plots are wavenumber low-pass and bottom plots, wavenumber high-pass (a) Strain values. (b) $f$-$k$ power spectrum. (c) 2-D autocorrelation function. }
    \label{fig:wavelet}
\end{figure*}

DAS signals exhibit strong temporal and spatial correlations that can be leveraged to improve compression. One effective method for exploiting these redundancies is linear predictive coding (LPC), which is conceptually equivalent to a linear prediction-error filter. In LPC, each signal is decomposed into two components: a predictable part that can be approximated as a linear combination of past values, and a residual part that captures the unpredicted variations. This decomposition is reversible via integer arithmetic, making it suitable for lossless compression. While it does not attempt to model the full physical complexity of DAS signals, it provides a compact representation that enhances compressibility. Specifically, for each signal \(c_i(t)\) we define:
\begin{equation}
\label{eq:lpc}
c_i(t) \;=\;\sum_{k=1}^{K} a_k\,c_i\bigl(t - k\bigr).
\end{equation}

The coefficients \(\{a_1, a_2, \dots, a_K\}\) are chosen to minimise the mean squared error between the actual sample \(c_i(t)\) and its linear prediction. We compute the optimal coefficients using the Levinson-Durbin recursion algorithm \citep{durbin60}, which leverages the Toeplitz form of the autocorrelation matrix estimated using the unbiased correlation estimator. The mean removal performed in the preprocessing step improves the model's fit by eliminating low-frequency bias in the signal. In practice, a separate LPC filter is first fitted and applied to each channel along the time dimension; the same procedure is then applied along the spatial dimension to the resulting residuals.

Fig. \ref{fig_lpc} depicts the effect of LPC on DAS data, using the example event from Vinaroz presented in section \ref{sec:pipe}. The top row shows the raw strain data, its frequency-wavenumber ($f$-$k$) transform, and its autocorrelation function. The bottom row shows the corresponding outputs after LPC filtering. By comparing the plots in panel (i), we observe how large-amplitude, long-period components are captured by the linear filter. In panel (ii) the low-frequency features are largely removed, and in panel (iii), the decorrelation effect is evident, with the autocorrelation significantly reduced in the LPC-filtered signal.

To demonstrate the effect of LPC within the pipeline, Fig. \ref{fig:wavelet} shows the four wavelet subbands for the same event, obtained after applying the wavelet transform in both the time and space dimensions, followed by LPC filtering. Notably, the low-frequency subbands exhibit stronger correlation, while the high-frequency components appear more uniform and less correlated.

Surface gravity-wave signals observed in shallow ocean-bottom DAS recordings are typically very strong (see, e.g., \citeauthor{romanowicz2023}, \citeyear{romanowicz2023}; \citeauthor{shi2025}, \citeyear{shi2025}), and can even obscure signals from nearby moderate earthquakes. However, the statistical properties of these waves tend to evolve slowly over time, and their velocity dispersion, under the linear approximation, is primarily controlled by ocean depth. These characteristics allow LPC to effectively track the signal evolution using only a few coefficients, significantly reducing the residual energy that needs to be stored after prediction.

\begin{figure}
    \centering
\includegraphics[width=0.5\linewidth]{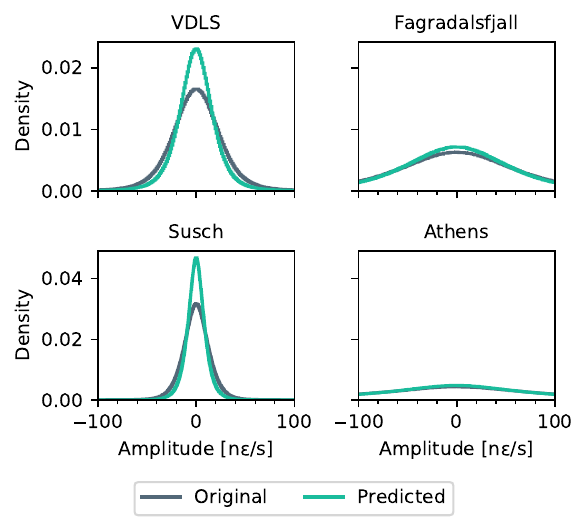}
    \caption{Density histograms of raw original, and predicted (wavelet-transformed and LPC-filtered) data, for datasets 1, 2, 5 and 8 (see Table \ref{tab:datasets}). The variance is reduced when the models are able to predict the signals. The lower the variance, the more compressible the data source is. The lower variance in the predicted data reveals how the processing contributes to better compression. }
    \label{fig_densities}
\end{figure}


\subsection{Entropy Coding}

Entropy coding is a lossless data compression technique that removes redundancy from a signal by assigning shorter code-words to more probable symbols. It can be applied to any stream of data, regardless of its origin. The goal is to approach the theoretical limit of compressibility as defined by the Shannon's source coding theorem \citep{shannon48}, which relates directly to the entropy of the signal. Entropy coding is a core component of high-performing lossless and lossy compression schemes. 

A fundamental concept in entropy coding is variable-length coding, where symbols are represented by bit-strings of varying lengths. By assigning shorter codewords to more frequently occurring symbols (and longer codewords to less frequent ones), a more efficient coding (i.e., better compression) can be achieved.

A widely used method for entropy coding is arithmetic coding \citep{witten87}. Arithmetic coding encodes the entire message as a single number between 0 and 1, allowing fine granularity in probability modelling and achieving near-optimal compression for signals with fixed distributions.

Examining the residuals obtained through our processing reveals a Gaussian-like distribution (see Fig. \ref{fig_densities}). For a (finely) quantised Gaussian random variable with variance $\sigma^2$, its entropy can be approximated by the differential entropy  $ H \approx \frac{1}{2} \log \bigl(2\pi e\,\sigma^2\bigr)$ \citep{cover2005}. Fig. \ref{fig_densities} also shows how the distribution of LPC-filtered data exhibit lower variance in several example datasets, effectively improving compression performance. 

DASPack leverages an existing implementation of arithmetic coding by \citet{bamler2022constriction}. A Gaussian distribution is fit for each one-dimensional channel, and the corresponding mean and variance are stored as parameters. To efficiently encode tail values and outliers, we use $k$-exponential Golomb codes, which provide a compact, variable-length representation for large integers \citep{Richardson2010}.




\section{Results}

\subsection{Compression efficiency and speed}
\label{sec:comp_eff}

We evaluate different compression algorithms by measuring the processing time and the compressed file size over 5-min excerpts randomly selected within each of the 15 datasets in Table \ref{tab:datasets}. We repeat this process for different quantisation steps, ranging from lossless mode to moderate degradation. 

Compression efficiency is evaluated in terms of the compression factor, defined as the ratio: $c_f = \frac{\text{Original Size}}{\text{Compressed Size}}$. This metric indicates the number of times the compressed dataset fits within the original size. It can also be interpreted as a measure of storage savings and as a proxy for file transfer time speed-up.

\begin{figure*}
    \centering
    \includegraphics[width=\linewidth]{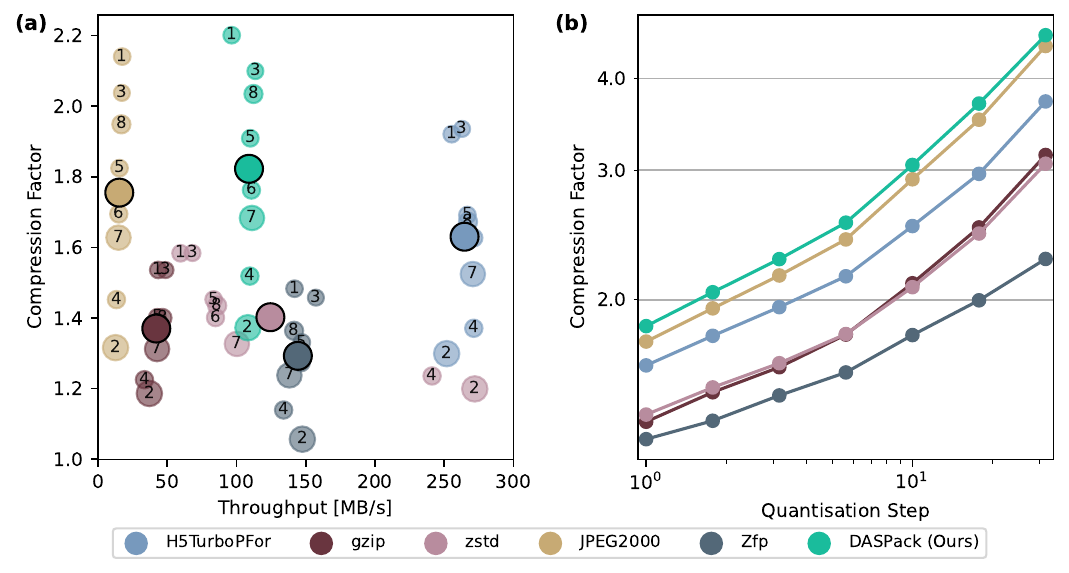}
    \caption{Compression factor of different methods for strain rate datasets (1-8). (a) Compression against throughput at lossless compression mode. The size of the points indicates the variance of the numbered dataset. The circle without a number is the centroid of the points, which is plotted to illustrate the trends. (b) Average compression across datasets at different quantisation steps. Our method, DASPack, beats all other methods in compression efficiency for all datasets, and competes in compression speed with the fastest methods.  }
    \label{fig_comp_strainrate}
\end{figure*}

\begin{figure*}
    \centering
    \includegraphics[width=\linewidth]{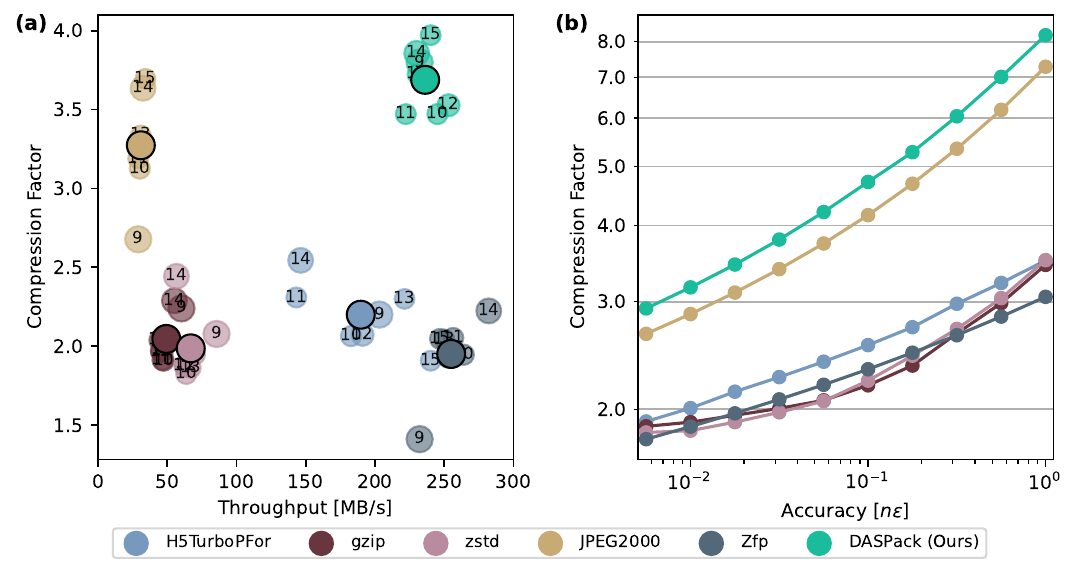}
    \caption{Compression factor of different methods for strain datasets (9-15). (a) Compression against throughput at a quantisation step $ \Delta = 10^{-1.5} ~\text{n}\varepsilon$. The size of the points indicates the variance of the numbered dataset. The circle without a number is the centroid of the points, which is plotted to illustrate the trends. (b) Average compression across datasets at different nanostrain accuracies. Our method, DASPack, beats all other methods in compression efficiency for all datasets, and is also the fastest in compression speed for strain data. }
    \label{fig_comp_strain}
\end{figure*}

Figs. \ref{fig_comp_strainrate} and \ref{fig_comp_strain} present a comparative analysis of the trade-off between compression speed (throughput in MB s\(^{-1}\) ) and efficiency (compression factor) for five compression methods (gzip, zstd, JPEG2000, H5TurboPFor, Zfp and DASPack) at different accuracies. Experiments were run in a conventional desktop computer (Apple M1 Pro CPU) in single-thread mode. Fig. \ref{fig_comp_strainrate} focuses on strain rate datasets 1-8, all recorded as 16-bit integers, while Fig. \ref{fig_comp_strain} focuses on strain datasets (9-15), recorded as 32-bit integers or floating-points. Because H5TurboPFor only works on 16-bit data, 32-bit integers are interpreted as two 16-bit values for this particular method. We fixed DASPack's parameters to square tiles of 2000 samples per side and two LPC coefficients, and implemented the steps using 32-bit integer operations. Strain data generally spans a higher dynamic range than strain rate due to the presence of high-amplitude, long period signals, hence requiring more bits to store raw data. Yet, for this same reason, strain data exhibit higher temporal correlation, hence achieving higher compression ratios.

Figures \ref{fig_comp_strainrate}a and \ref{fig_comp_strain}a reveal the performance of different methods for each individual dataset at a given fixed-accuracy (lossless for strain rate, $\Delta = 10^{-1.5}~\text{n}\varepsilon$ for strain). Overall, gzip and zstd achieve moderate compression factors but exhibit relatively low throughput, generally below 100 MB/s. H5TurboPFor and Zfp prioritise speed, often exceeding 200 MB/s, and H5TurboPFor maintains high compression factors for strain rate data. Zfp performs poorly in strain rate because it is a method optimised for floating-points; but the performance is only moderately better for strain data. JPEG2000 achieves higher compression factor, but is the slowest method, achieving 10-20~MB~s\(^{-1}\). DASPack, by contrast, offers both high throughput (up to 250~MB~s\(^{-1}\) for strain, 125~MB~s\(^{-1}\) for strain rate) and the strongest compression performance. It is important to note that compression factors are dataset-dependent and should not be taken as absolute numbers. While different datasets may yield higher or lower compression factors, the consistent trend is that DASPack outperforms alternative methods across all cases considered. DASPack consistently occupies a favourable region of the plot, offering high compression ratios at speeds that match or exceed leading alternatives. This makes it a strong choice for applications requiring both data reduction and fast, scalable performance.

Figures \ref{fig_comp_strainrate}b and \ref{fig_comp_strain}b demonstrate how the compression factor evolves as accuracy constraints (the quantisation step, in nano-strain rate and nano-strain units) are relaxed. As expected, all methods achieve increasingly higher compression factors as the allowed tolerance increases, since relaxed accuracy requirements directly translate into smaller compressed file sizes. For strain datasets, the compression is shown for varying accuracy in nanostrain units. In contrast, for strain rate data, the effective resolution varies with the sampling rate: since strain rate is computed by dividing phase change by the sampling interval, the quantisation step in strain rate increases with sampling rate. As a result, higher sampling rates yield coarser strain rate resolution. To allow consistent comparison across strain rate datasets, we express the quantisation step relative to the raw 16-bit integer values, where a step of 1 corresponds to lossless compression and a step of 10 represents a one-digit reduction in absolute precision.

JPEG2000 and DASPack provide consistently superior compression performance at higher tolerances, reaching compression factors beyond $8\times$. Even more relaxed accuracies (such as $1  ~\text{n}\varepsilon$ or a $10^2$ quantisation step) can result in up to $10\times$ compression for some datasets. In certain applications or datasets, this may offer a viable trade-off without compromising critical signal content.  

The results indicate that DASPack’s fixed-accuracy compression approach is not only able to effectively leverage lossy compression, but also highly competitive when precision constraints are relaxed, making it a robust choice for applications that require significant data reduction with controlled accuracy.

\subsection{Application examples}

DASPack is suited to every DAS application by appropriately tuning the quantisation step to balance data fidelity and compression efficiency. To demonstrate its performance, this section presents several examples of compressed real events with varying levels of degradation. Each example also includes error maps, defined as the difference between the original and compressed signals.

\begin{figure*}
    \centering
    \includegraphics[width=\linewidth]{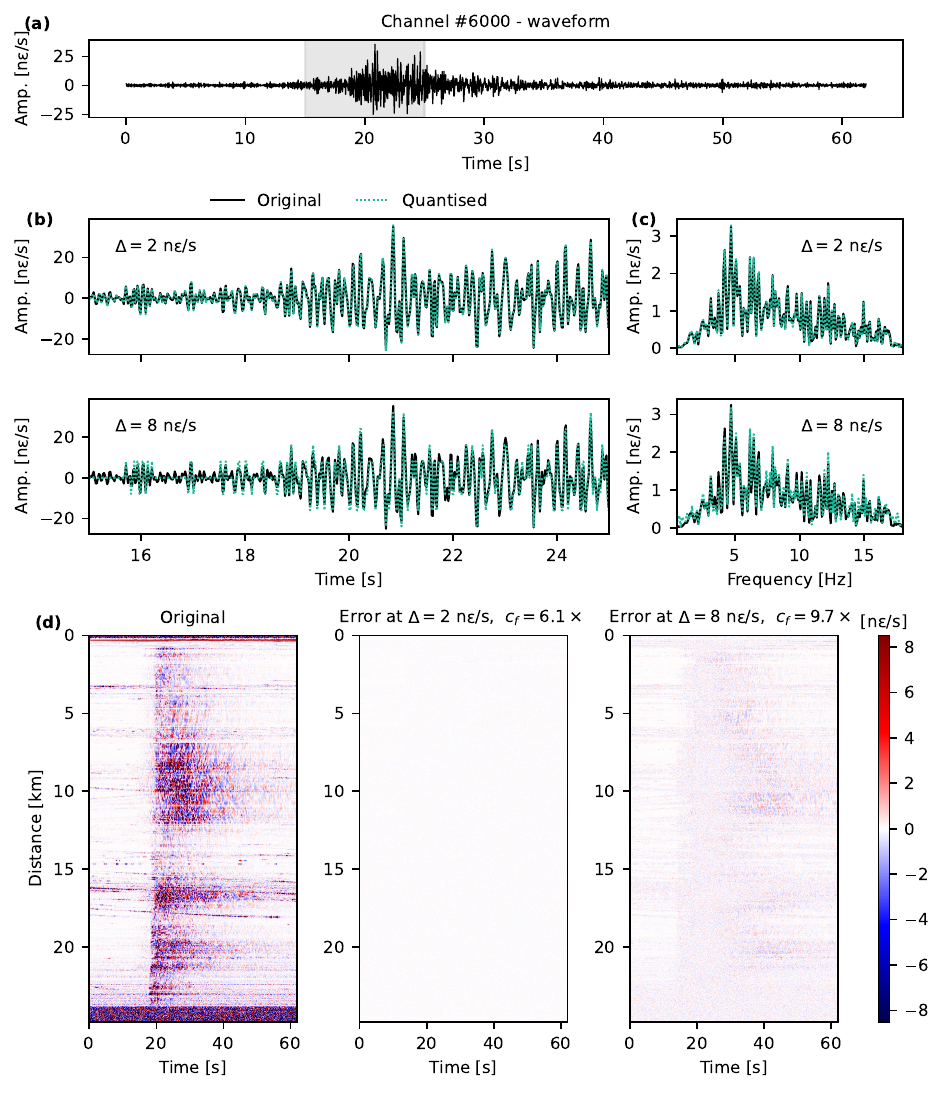}
    \caption{A regional earthquake of magnitude \textit{ML} 3.4 occurred 48 km northwest of the interrogator on 23 September 2021 at 07:06:54 UTC during the Athens experiment (dataset 2). The figure compares unfiltered raw signals with quantised versions using two degradation levels ($\Delta = 2 ~\text{n}\varepsilon~\text{s}^{-1}$ and $\Delta = 8 ~\text{n}\varepsilon~\text{s}^{-1}$), corresponding to compression factors of $c_f = 6.1\times$ and $c_f = 9.7\times$. (a) Waveform from channel 6000 (at 6000 m). (b) Zoomed view of the waveform. (c) Discrete Fourier Transform of the zoomed window. (d) 2-D strain map of the original signal (left) and error maps between the original and degraded signals (middle and right). }
\label{fig_athens_experiment}
\end{figure*}

Fig. \ref{fig_athens_experiment} displays unfiltered strain rate data from an \textit{ML} 3.4 regional earthquake recorded in Athens (dataset 2), with a dynamic range of ±25 ~n$\varepsilon~\text{s}^{-1}$. Two quantisation steps are applied: $\Delta = 2 ~\text{n}\varepsilon~\text{s}^{-1}$ and $\Delta = 8 ~\text{n}\varepsilon~\text{s}^{-1}$. Results show that with low step values, the quantised trace is virtually indistinguishable from the original, indicating negligible error. In contrast, at high quantisation levels, low-amplitude features disappear, and only the largest peaks remain, indicating information loss, a trend that is likewise observed in the frequency domain. This is confirmed by the error map, which  is nearly blank for $2 ~\text{n}\varepsilon~\text{s}^{-1}$, while the $8 ~ ~\text{n}\varepsilon~\text{s}^{-1}$ case reveals systematic residuals, especially for low-amplitude arrivals. Compression factors increase accordingly, from $6.1\times$ to $9.7\times$. The completely lossless mode achieves a compression factor of $2.8\times$ on the same data, demonstrating the importance of allowing slight data degradation.

\begin{figure*}
    \centering
    \includegraphics[width=\linewidth]{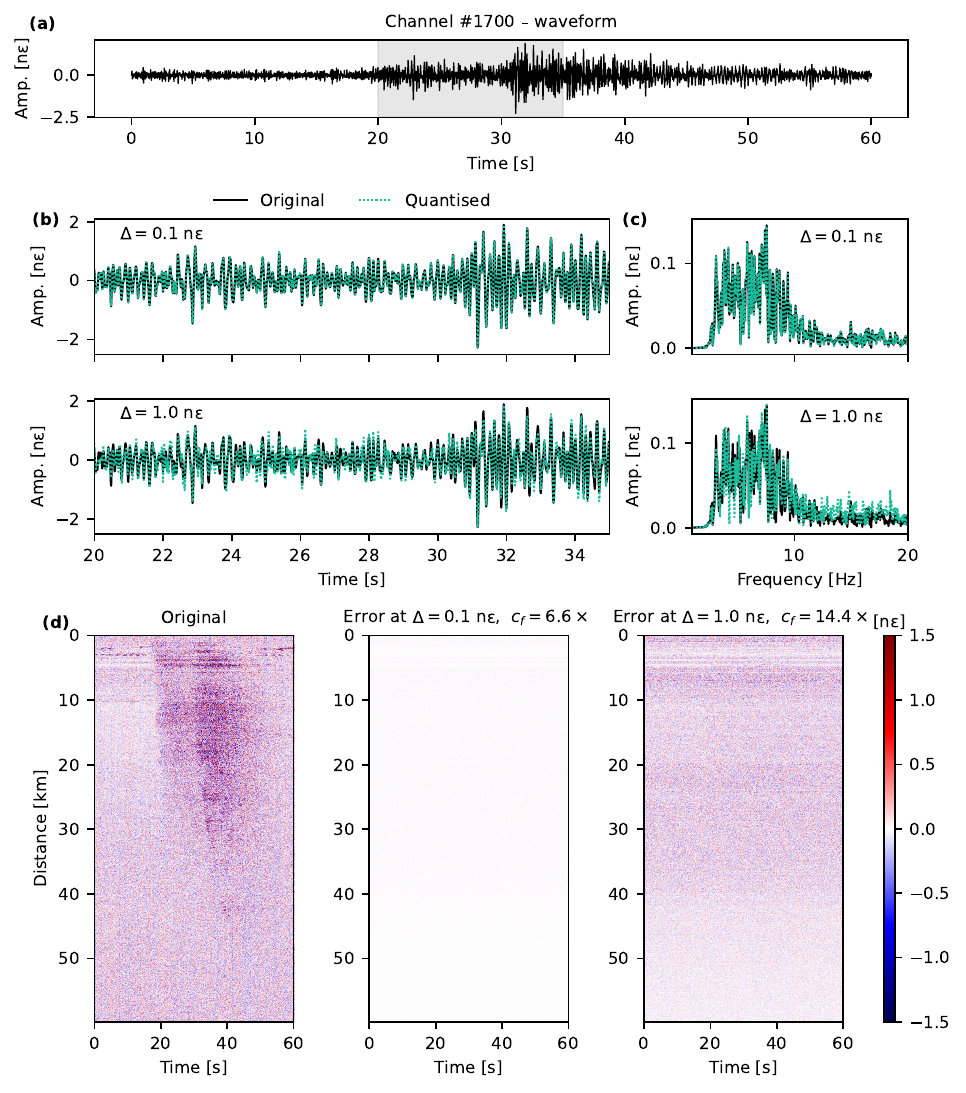}
    \caption{An onshore earthquake of magnitude \textit{mbLg} 3.1 occurred 97 km northeast of the interrogator on 5 October 2023 at 21:40:04 UTC and was recorded on a submarine fibre-optic cable, in the Alboran Sea (dataset 10). The figure compares filtered raw signals with quantised versions using two degradation levels ($\Delta = 0.1 ~\text{n}\varepsilon$ and $\Delta = 1 ~\text{n}\varepsilon$), corresponding to compression factors of $c_f = 6.6\times$ and $c_f = 14.4\times$. (a) Waveform from channel 1700 (at 17,000 m). (b) Zoomed view of the waveform. (c) Discrete Fourier Transform of the zoomed window. (d) 2-D strain map of the original signal (left) and error maps between the original and degraded signals (middle and right). }
\label{fig_safe_experiment}
\end{figure*}

 Fig. \ref{fig_safe_experiment} presents filtered strain data from a magnitude \textit{mbLg} 3.1 earthquake recorded in the Alboran Sea (dataset 10). This example illustrates how DASPack performs when the signal amplitude is very small (only about $\pm 2.5 ~\text{n}\varepsilon$ across the entire event). In this case, quantisation steps of  $\Delta = 0.1 ~\text{n}\varepsilon$ and $\Delta = 1 ~\text{n}\varepsilon$ result in compression factors of $6.6\times$ and $14.4\times$, respectively. At the finer step, the quantised waveform closely matches the original, with negligible error. At $1 ~\text{n}\varepsilon$, the error map shows residuals below $0.5 ~\text{n}\varepsilon$ that do not exhibit any visual patterns, indicating that most of the signal energy survives even at this aggressive compression level.

\begin{figure*}
    \centering
    \includegraphics[width=\linewidth]{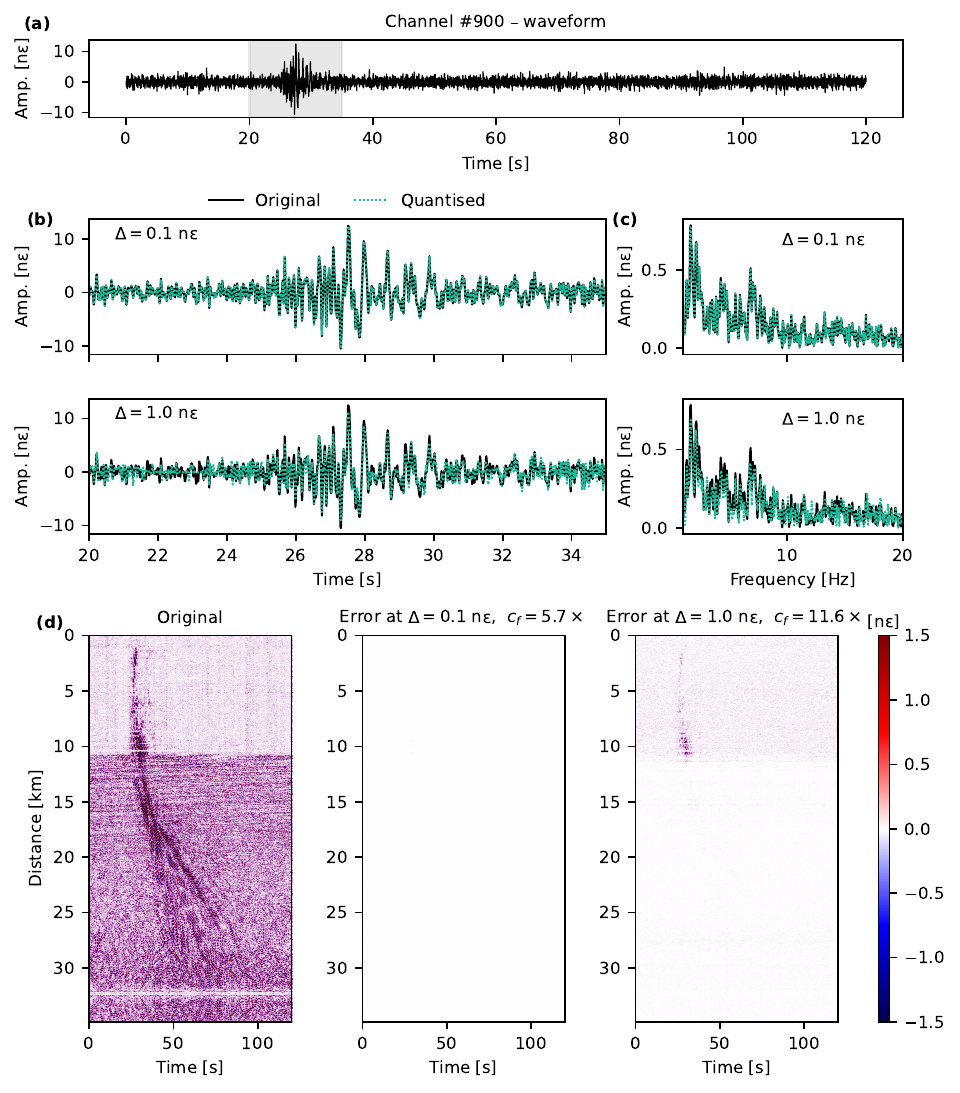}
    \caption{A quarry blast occurred 26 km southeast of the interrogator on 9 June 2023 at 10:26:30 UTC and was recorded during the Vinaroz experiment (dataset 13). The figure compares filtered raw signals with quantised versions using two degradation levels ($\Delta = 0.1  ~\text{n}\varepsilon$ and $\Delta = 1 ~\text{n}\varepsilon$), corresponding to compression factors of $c_f = 5.7\times$ and $c_f = 11.6\times$. (a) Waveform from channel 900 (at 9,000 m). (b) Zoomed view of the waveform. (c) Discrete Fourier Transform of the zoomed window. (d) 2-D strain map of the original signal (left) and error maps between the original and degraded signals (middle and right).}
\label{fig_castor_experiment}
\end{figure*}

Similar patterns are observed in Fig. \ref{fig_castor_experiment}, which shows filtered strain data from a quarry blast recorded during the Vinaroz experiment (dataset 13), with amplitudes reaching about $\pm 10 ~\text{n}\varepsilon$. Quantisation steps of $\Delta = 0.1 ~\text{n}\varepsilon$ and $\Delta = 1 ~\text{n}\varepsilon$ result in compression factors of $5.7\times$ and $11.6\times$, respectively. In this example, the error is not distributed uniformly: offshore channels (beyond 10 km) retain nearly all signal information even at the higher quantisation level, while the onshore section (within the first 10 km) records lower amplitudes and threfore exhibits higher residuals in the error maps.

\begin{figure*}
    \centering
    \includegraphics[width=\linewidth]{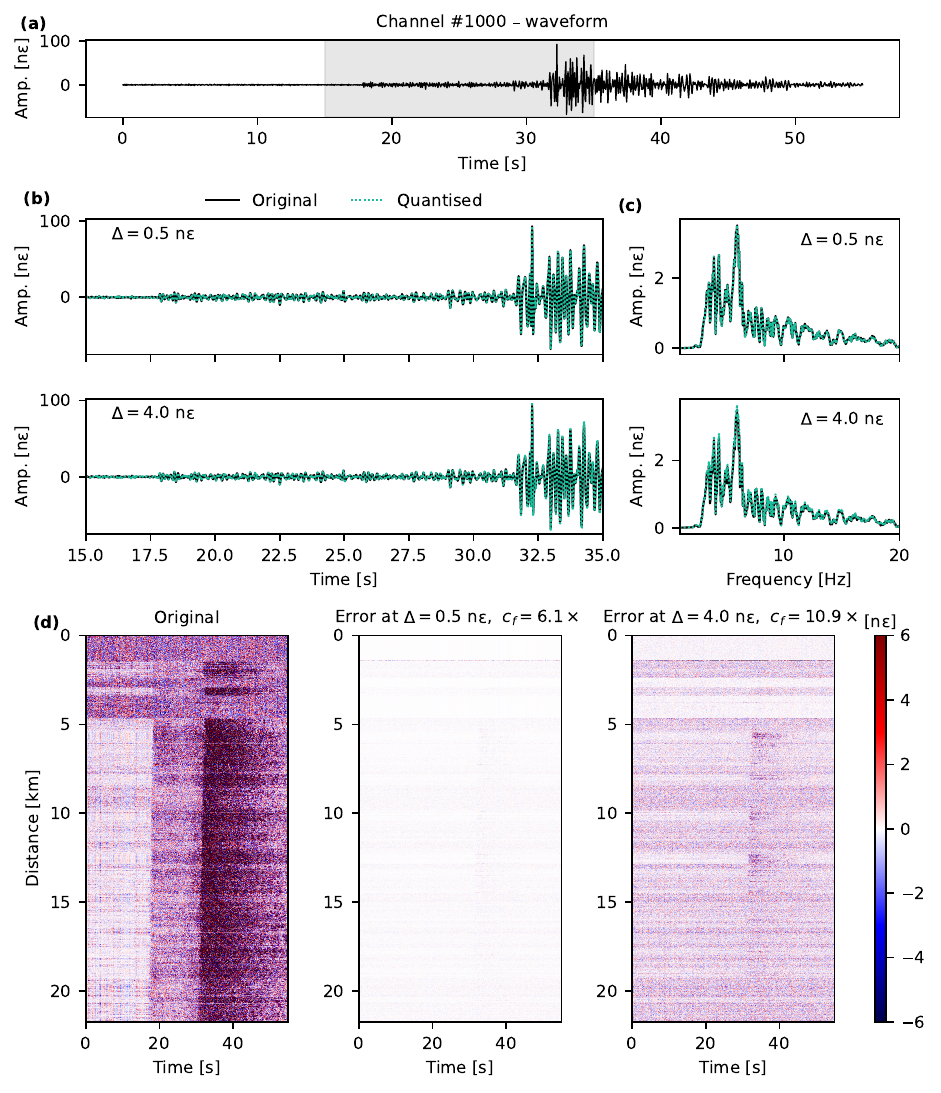}
    \caption{An earthquake of magnitude \textit{mbLg} 4.9 occurred 148 km west from the interrogator on 26 October 2021 at 16:25:37 UTC during the La Palma volcanic eruption and was recorded at the Teide Astronomical Observatory (dataset 14). The figure compares filtered raw signals with quantised versions using two degradation levels ($\Delta = 0.1 ~\text{n}\varepsilon$ and $\Delta = 4 ~\text{n}\varepsilon$), corresponding to compression factors of $c_f = 6.1\times$ and $c_f = 10.9\times$. (a) Waveform from channel 1000 (at 10,000 m). (b) Zoomed view of the waveform. (c) Discrete Fourier Transform of the zoomed window. (d) 2-D strain map of the original signal (left) and error maps between the original and degraded signals (middle and right).
    }
\label{fig_tenerife_experiment}
\end{figure*}

Finally, Fig. \ref{fig_tenerife_experiment} extends the examples to a high-amplitude ($\pm 100 ~\text{n}\varepsilon$), 4.9 \textit{mbLg} earthquake recorded during the 2021 La Palma eruption (dataset 15). This case confirms that DASPack maintains high fidelity at moderate quantisation. For example,  $\Delta = 4 ~\text{n}\varepsilon$  results in a compression factor of $10.9\times$. In this higher degradation case, the error appears to be mostly uncorrelated, with higher values along the high-amplitude front.

\section{Discussion}

\subsection{Compression efficiency}

DASPack contributes to DAS data management by providing sustained, real-time throughput while simultaneously delivering the highest compression factors across a broad range of acquisition settings. Its performance advantage is most pronounced for strain data, where the combination of wavelet decomposition and linear predictive coding (LPC) removes the slowly varying, large-amplitude components that dominate the signal variance. Because strain is typically stored in 32-bit words to accommodate this broad dynamic range, both the entropy reduction and the arithmetic-coding efficiency are amplified, yielding compression factors that exceed $8\times$. strain rate, in contrast, is acquired with narrower dynamic range and usually quantised at 16 bits; here the algorithm still outperforms specialised competitors with smaller margins, confirming that the same pipeline remains effective even when the predictable, low-frequency portion of the spectrum is reduced. This difference in performance arises not from the compression method itself but from the statistical properties of the input data: as a temporal derivative of strain, strain rate naturally exhibits lower amplitude variability and reduced temporal coherence, which limits the predictive power of both wavelet and LPC stages and reduces overall compressibility.

A central design choice in DASPack is the explicit separation of the scientific decision from engineering optimisation. By restricting every lossy decision into a single, user-controlled quantisation step, the method offers a guarantee that the reconstruction error is strictly bounded and readily interpretable as an absolute amplitude. This contrasts with perceptual or signal-adaptive codecs in which loss is distributed across many internal stages and is therefore difficult to predict. The fixed-accuracy paradigm enables practitioners to tailor the degradation to the specific use case: a researcher interested in microseismic events may keep the step below $0.05 ~\text{n}\varepsilon$ to preserve the smallest arrivals, whereas monitoring traffic-induced vibration along a bridge may tolerate $0.1~\text{n}\varepsilon$ and enjoy a four-fold reduction in storage cost.

\subsection{Compression speed and real-time capability}

\begin{figure}
    \centering
    \includegraphics[width=0.6\linewidth]{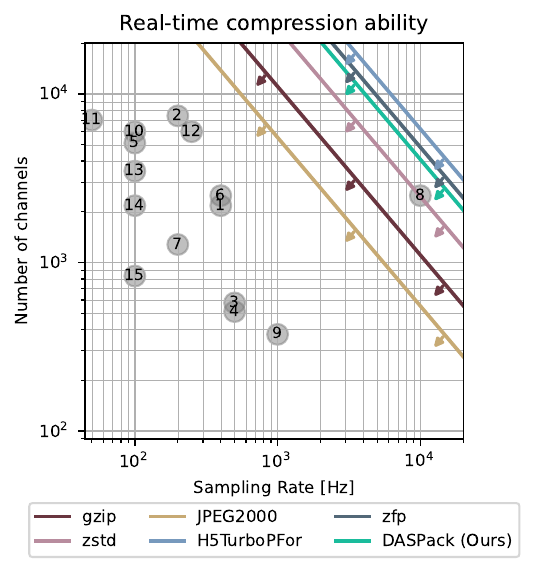}
    \caption{ Real-time compression envelope. 
    Each marker corresponds to one of the 15 datasets (Table \ref{tab:datasets}) and is positioned by its number of channels (x-axis) and sampling rate (y-axis). Solid curves are the real-time limits of five compression methods: for any point below a curve the method can keep up with the raw data stream, because $R \leq T_{\text{method}}$ (see equation \ref{eq:throughput}). Datasets lying above a curve would overwhelm that compressor. }
    \label{fig_throughput}
\end{figure}

For a given interrogator, the raw data generation rate is
\begin{equation}
R \;=\; N_\mathrm{ch}\,f_s\,b
\quad[\mathrm{bytes\,s^{-1}}],
\label{eq:throughput}
\end{equation}

where  
$N_\mathrm{ch}$ is the number of channels,  
$f_s$ the sampling rate (samples s\(^{-1}\)), and  
$b$ the datum size in bytes (2 for int16, 4 for int32 and float32). A compressor with sustained throughput \(T_\text{method}\) can operate in real time whenever \(R \le T_\text{method}\).  
We therefore plot in Fig. \ref{fig_throughput} the level curves \(R = T_\text{method}\) for each algorithm, using the average \(T_\text{method}\) values measured in section \ref{sec:comp_eff}. The visual test is simple: if a point falls below a curve, that algorithm can compress the dataset on-the-fly. Otherwise, the raw stream outruns the compressor.

All datasets fall below the DASPack curve, confirming that DASPack can handle even the most demanding case: Zurisee (dataset 8), which involved an underwater application with a 10 kHz sampling rate across 2,496 channels, aimed at recording human and animal activity. JPEG2000, Zstd and Gzip fail for this extreme. DASPack, Zfp and H5TurboPfor are observed to be able to compress all configurations in real time.

In practice, DASPack uses 32-bit integer operations and achieves a sustained compression speed of approximately 250~MB~s\(^{-1}\) for strain data when running in single-threaded mode. For strain rate data, typically stored as 16-bit integers, performance is about half, around 125~MB~s\(^{-1}\). Thanks to independent per-tile processing, DASPack supports parallelism. When configured with 8 threads on an Apple M1 Pro CPU, the average compression speed across our datasets increases to roughly 750~MB~s\(^{-1}\). This throughput exceeds the write speed of standard hard disk drives (HDDs), which typically peak around 200~MB~s\(^{-1}\), making storage, rather than computing, the limiting factor in many setups. To fully leverage DASPack’s compression speed, faster storage such as solid-state drives (SSDs), which can sustain several gigabytes per second, is recommended. Clusters with a large number of nodes and drive bays offering high read and write speeds further exploit parallelism, enabling highly efficient compression and decompression.

\subsection{Understanding the effect of quantisation}

The controlled degradation step in DASPack is based on fixed-accuracy quantisation, which has well-understood statistical properties. For sufficiently small quantisation steps, the rounding error behaves like additive uniform noise with zero mean and known variance \citep{Widrow_2008}. This behaviour explains why the resulting error is often modeled as white noise, and was experimentally observed in Figures \ref{fig_athens_experiment}--\ref{fig_tenerife_experiment}b,d. However, unlike truly independent noise, quantisation can lead to repeated values in low-amplitude segments. For example, entire portions of a trace being rounded to zero. At very coarse quantisation levels, this may introduce visible artifacts, even if the overall frequency content remains largely preserved.

Figures \ref{fig_athens_experiment}--\ref{fig_tenerife_experiment}c illustrate that the frequency spectrum is barely modified by quantisation, and only the very low-amplitude parts of the signal are suppressed. Indeed, interpreted as random noise, the quantisation error converges to Gaussian behaviour in the frequency domain \citep{oppenheim2009}.

\begin{figure*}
    \includegraphics[width=0.95\linewidth]{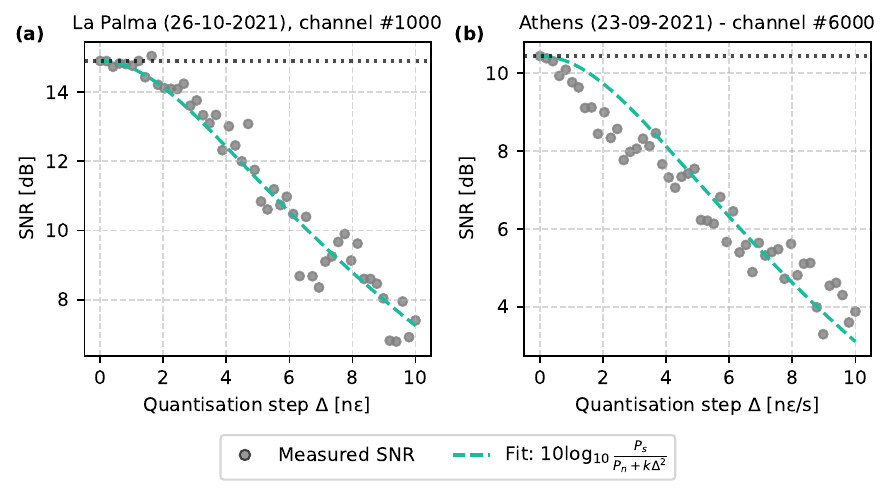}
    \caption{Signal-to-noise ratio (SNR) as a function of quantisation step $\Delta$ for two DAS datasets. 
    (a) La Palma eruption (26 October 2021, channel \#1000), also shown in Fig. \ref{fig_tenerife_experiment} and (b) Athens array (23 September 2021, channel \#6000), also shown in Fig. \ref{fig_athens_experiment}. 
    Grey markers denote measured SNR values, while dashed green lines show fitted theoretical predictions of the form 
    $10\log_{10}\frac{P_s}{P_n+k\Delta^2}$. The fitted constants are $k=0.013$ ($R^2=0.95$) for La Palma and $k=0.034$ ($R^2=0.87$) for Athens. 
    For comparison, ideal i.i.d.\ uniform quantisation noise would correspond to $k=1/12\approx0.083$. 
    Dithering was applied during the SNR calculation to mitigate artificial vanishing errors at large $\Delta$.}
    \label{fig_snr_vs_delta}
\end{figure*}

Another perspective on quantisation is to view it as a uniform degradation of the signal-to-noise ratio (SNR). If the error behaves like additive uniform noise, its power increases proportionally to the square of the quantisation step \citep{Widrow_2008}. This is consistent with the results in Figure \ref{fig_snr_vs_delta}, where the SNR was computed as the ratio of signal and noise window powers. To prevent spurious vanishing errors caused by large quantisation steps rounding entire traces to zero, dithering was added during the SNR calculation, which helps recover idealised behaviour of quantisation noise \citep{Widrow_2008}. Importantly, dithering was introduced solely during SNR evaluation to obtain meaningful estimates at coarse quantisation levels; it is not part of the DASPack method itself. The measured SNR follows closely the expected noise law, providing an additional validation of the theoretical model. This analysis demonstrates that quantisation in DASPack is not only interpretable as a bounded reconstruction error, but can also be understood within the familiar framework of additive noise, with clear implications for the effective resolution and dynamic range of compressed DAS signals.

While the statistical interpretation of quantisation is useful, it does not by itself capture the full implications for downstream seismic analysis. DAS is applied to a wide variety of problems, each of which may respond differently to small compression errors. For instance, typical processing tasks include (i) detecting an event through STA/LTA or template matching, (ii) picking traveltimes using manual, automatic, or array-based approaches, (iii) estimating dispersion curves from ambient noise correlations, or (iv) measuring amplitude decay to infer attenuation. Each of these workflows involves distinct sensitivities to low-amplitude signals, phase accuracy, or spectral content. A superficial treatment of all these aspects would risk being incomplete or even misleading. For this reason, we deliberately refrain from providing application-specific guidelines in this paper. Instead, we emphasise that a rigorous evaluation of the effect of controlled degradation on the accuracy of diverse seismological analyses deserves a dedicated study.

\subsection{Analysis of DASPack's lossless pipeline}

The wavelet front-end contributes primarily by aligning the data with the locality assumption of LPC. DAS records exhibit strong temporal correlation that is well captured by low-order predictors, especially after the spatial and temporal low-pass information has been separated from the high-pass detail. The 5/3 bi-orthogonal wavelet was selected because it allows an integer-to-integer lifting implementation, guaranteeing perfect reversibility. Although cascading the transform in multiple steps was tested, a single step already removed most inter-sample redundancy; additional levels did not improve compression efficiency.

Linear predictive coding plays a complementary role by flattening the residual spectrum. Figures \ref{fig_lpc} and \ref{fig:wavelet} show that, after LPC, the residuals better approximate white noise, a condition under which arithmetic coding approaches the Shannon limit. Prior to this, subtracting the temporal mean (stored as side information) was found to improve stationarity and thus yield a more effective model. In contrast, removing the spatial mean did not provide any noticeable gain in compression. The order of the predictor is kept small (typically two or three in time and one in space) to bound the side information that must be stored with each tile.  Larger numbers of coefficients were not found to improve compression efficiency and, in fact, tend to degrade it due to the increased volume of side information. Since these coefficients are stored independently for each tile, smaller tile sizes lead to more frequent coefficient storage, which increases overhead and hinders overall compression. However, excessively large tiles may span signals with varying statistical properties, reducing the effectiveness of the linear prediction. A balance between these effects was found by using square tiles of 1000 or 2000 samples per side, but the optimal choice depends on the data.

Beyond raw performance, DASPack addresses practicalities that often hinder adoption of research prototypes. The core is written in Rust, a language that combines speed with memory-safety guarantees. The code interface is designed to integrate seamlessly with existing seismic workflows and can be combined with widely used formats such as HDF5 and Zarr. Python bindings expose the compressor directly to popular data-analysis environments, so that researchers can read and write compressed files with little effort. Default parameters were established for the processing steps, accommodating the vast majority of cases and resulting in only marginal variations in compression efficiency, conveniently relieving users from the need to adjust them.

\subsection{Combination with sampling methods}

Beyond entropy-based compression, decimation and subsampling are widely used signal processing techniques that reduce data volume by lowering the temporal or spatial sampling rate. These operations can be combined with DASPack to achieve further reductions in storage requirements, provided that the reduced sampling rate preserves the bandwidth of interest. A more advanced alternative is compressed sensing (CS), which exploits the sparsity of many DAS wavefields. By leveraging this sparsity, CS allows signals to be sampled at rates below the Nyquist limit while still enabling accurate reconstruction. This makes CS particularly well-suited for application-specific DAS deployments where only a subset of information (e.g., source localisation or detection of transient events) is required. For example, \citet{Wang25} presented a DAS system achieving a 40-fold lossy data reduction by combining envelope extraction with CS, while preserving accurate event localisation. Although CS is not universally applicable due to its reliance on signal sparsity and the complexity of reconstruction, it is a powerful tool in targeted deployments. Importantly, DASPack is complementary to such approaches: it can be applied downstream in lossless mode after CS or other degradation techniques, providing additional entropy-based compression without further information loss. This integration is especially attractive for environments where storage and transmission resources are constrained.

\subsection{Comparison and potential combination with deep learning methods}

Recent studies have proposed deep learning for DAS data reduction, primarily through subsampling with reconstruction \citep{yiyu2024} and transformer-based autoencoders \citep{chen2025}. While effective in extracting higher-order structure, these approaches fundamentally differ from DASPack. \citet{yiyu2024} approximate missing traces from downsampled data, whereas \citet{chen2025} encode DAS tiles into latent features, storing decoder weights for signal regeneration. Both strategies produce approximations without explicit error bounds, and may obscure low-amplitude or transient events. In contrast, DASPack provides deterministic guarantees: either exact lossless or fixed-accuracy lossy compression with strictly bounded error.

The computational trade-offs are also substantial. \citet{chen2025} report $\sim 30$ minutes of GPU training for a one-minute DAS segment, with transfer learning partially mitigating the cost. \citet{yiyu2024} achieved 51 seconds of A100 GPU compute to reconstruct a 10-minute record, and $\sim 80$ reconstructions per second on a Raspberry Pi for 1,000 channels. This is equivalent to a $\sim 3\times$ real-time throughput at 25 Hz sampling, with poor scaling for larger deployments. In comparison, DASPack compresses and decompresses in real time on standard CPUs, achieving sustained throughput up to 200 MB~$s^{-1}$ using one thread, without requiring GPUs or model storage. This is equivalent to real-time sampling throughput of about 50,000 Hz for the same example of 1,000 channels. 

Importantly, downsampling or approximation should not be confused with entropy-based compression. Subsampling discards samples and interpolates missing information, while DASPack re-encodes the full signal bandwidth with user-defined error control. This distinction implies complementarity: DASPack can be applied prior to or after learned reconstruction, providing reliable entropy reduction while leaving room for interpolation-based inpainting. Likewise, DASPack’s controlled quantisation could act as a regulariser for neural models, suppressing noise below application thresholds and potentially accelerating training convergence.

\subsection{Applications of DASPack}

Applications of DASPack extend across the spectrum of DAS research and industry practice. In earthquake seismology, the ability to record multi-month campaigns with significantly reduced storage opens the door to monitoring arrays spanning hundreds of kilometres of fibre with reduced disk budgets.  Environmental sensing applications, such as glacier dynamics, landslide detection or traffic-induced urban vibration benefit similarly because the relevant signals can sustain different levels of quantisation without compromising scientific utility.

The fixed-accuracy compression approach allows practitioners to choose a quantisation step that is safely below the minimum signal amplitude of interest. Figs. \ref{fig_athens_experiment}, \ref{fig_safe_experiment}, \ref{fig_castor_experiment} and \ref{fig_tenerife_experiment} demonstrate this principle: near-perfect signal fidelity is achieved at moderate quantisation steps, achieving up to $6\times$ compression factors.

Controlled compression is well suited to domains that produce large data volumes where only a small fraction contains transient events. With automatic event detection and phase picking, adaptive compression strategies can be applied: data windows with strong signals are stored at high fidelity, while quieter periods are compressed more aggressively. This approach can reduce storage use without compromising the integrity of relevant data.

Finally, because of its high throughput and low hardware requirements, DASPack can be deployed directly on edge devices or interrogators, enabling real-time compression of DAS data at the source before transmission or storage. This is particularly valuable for remote or bandwidth-limited deployments, where GSM, satellite or intermittent connectivity makes raw data handling challenging.

\section{Conclusions}
We have presented DASPack, a high-performance, open-source compression tool specifically tailored for DAS data. DASPack addresses the growing challenge of managing the massive data volumes generated by DAS deployments, offering a unique combination of controlled, user-defined degradation and an efficient, fully reversible lossless compression pipeline. Its fixed-accuracy quantisation approach allows users to make informed decisions about data fidelity while ensuring optimal compression.

Across a diverse set of 15 real-world DAS datasets, DASPack consistently outperformed both general-purpose and domain-specific compression algorithms in terms of compression efficiency and processing speed. File size reductions of up to $3\times$ were achieved in lossless mode, with up to $6\times$ compression realised with minimal information loss, and $10\times$ compression attained with acceptable degradation. It demonstrated real-time capability even under demanding conditions, such as high channel counts and high sampling rates. The tool supports a broad range of acquisition scenarios and integrates easily into existing workflows via HDF5 support and Python bindings.

DASPack also enables efficient on-site compression, making it suitable for deployment in remote or bandwidth-limited environments. Its principled design empowers users to balance storage savings with scientific accuracy, providing a practical and scalable solution for long-term DAS data management. We believe DASPack can play an important role in future DAS deployments, supporting applications across seismology, environmental sensing, infrastructure monitoring, and beyond.


\begin{acknowledgments}

We thank Professor Francesco Grigoli and an anonymous reviewer for the valuable feedback that helped improve the manuscript. 

The first author thanks Professor Cesar R. Ranero for his support during the realisation of the work.

This work was supported in part by the "Severo Ochoa Centre of Excellence" program (CEX2019-000928-S-21-2); the European Union NextGenerationEU/PRTR Program under projects PSI (PLEC2021-007875), TREMORS (CPP2021-008869), the CSIC contract SAFE with Telxius Cable España, S. L. (ref. 20233069) and the Spanish Ministry of Science, Innovation and Universities under DeeLight project (PID2020-117142GB-I00).

Canalink, IslaLink, Telxius, ENAGAS, Teide and Roque de los Muchachos Astronomical Observatories facilitated fibre-optic cable access. Aragon Photonics provided the HDAS interrogator used in some of the experiments.
\end{acknowledgments}

\begin{dataavailability}
DASPack is openly available at \url{https://github.com/asleix/daspack}, including examples to use the software. We used an existing entropy coding implementation named Constriction, \url{https://github.com/bamler-lab/constriction}. For published datasets listed in Table \ref{tab:datasets}, details on open data access can be found in the correponding publications. Unpublished datasets are  available from the respective authors upon reasonable request. Data segments used in Figs. \ref{fig_athens_experiment}--\ref{fig_tenerife_experiment} are available at \url{https://doi.org/10.20350/digitalCSIC/17586} \citep{daspack_dataset}. 


\end{dataavailability}

\bibliographystyle{gji}
\bibliography{references.bib}




\label{lastpage}

\end{document}